\documentclass[conference]{IEEEtran}
\usepackage{graphicx}
\usepackage{amsfonts}
\usepackage{dblfloatfix}
\usepackage{array}
\ifCLASSOPTIONcompsoc
	\usepackage[caption=false,font=normalsize,labelfont=sf,textfont=sf]{subfig}
\else
	\usepackage[caption=false,font=footnotesize]{subfig}
\fi

\newcommand{\MYfooter}{\smash{
\hfil\parbox[t][\height][t]{\textwidth}{\centering
\thepage}\hfil\hbox{}}}
\makeatletter

\def\ps@headings{%
\def\@oddhead{\parbox[t][\height][t]{\textwidth}{\centering \normalsize{
2016 IEEE/ACM International Conference on Advances in Social Networks Analysis and Mining (ASONAM)}\\
\noindent\makebox[\linewidth]{\rule{\textwidth}{0pt}}
}\hfil\hbox{}}%

\def\@evenhead{\parbox[t][\height][t]{\textwidth}{\centering
2016 IEEE/ACM International Conference on Advances in Social Networks Analysis and Mining (ASONAM)\\
\noindent\makebox[\linewidth]{\rule{\textwidth}{0pt}}
}\hfil\hbox{}}%

\def\@oddfoot{\MYfooter}%
\def\@evenfoot{\MYfooter}}

\def\ps@IEEEtitlepagestyle{%
\def\@oddhead{\parbox[t][\height][t]{\textwidth}{\centering\normalsize{
2016 IEEE/ACM International Conference on Advances in Social Networks Analysis and Mining (ASONAM)}\\
\noindent\makebox[\linewidth]{\rule{\textwidth}{0pt}}
}\hfil\hbox{}}%
\def\@evenhead{\scriptsize\thepage \hfil \leftmark\mbox{}}%
\def\@evenfoot{\MYfooter}}

\DeclareRobustCommand*{\IEEEauthorrefmark}[1]{%
  \raisebox{0pt}[0pt][0pt]{\textsuperscript{\footnotesize #1}}%
}

\makeatother
\begin{document}
\title{A Comparative Study of Social Network Classifiers for Predicting Churn in the Telecommunication Industry}
\author{\IEEEauthorblockN{Mar\'{i}a \'{O}skarsd\'{o}ttir\IEEEauthorrefmark{1}, 
Cristi\'{a}n Bravo\IEEEauthorrefmark{2}, 
Wouter Verbeke\IEEEauthorrefmark{3}, 
Carlos Sarraute\IEEEauthorrefmark{4},
Bart Baesens\IEEEauthorrefmark{5}\IEEEauthorrefmark{6}
and Jan Vanthienen\IEEEauthorrefmark{6}}

\IEEEauthorblockA{\IEEEauthorrefmark{1}Department of Computer Science, Reykjavik University, Iceland,  Email: mariaoskars@ru.is}
\IEEEauthorblockA{\IEEEauthorrefmark{2}Department of Statistical and Actuarial Sciences, University of Western Ontario, Canada, Email: cbravoro@uwo.ca}
\IEEEauthorblockA{\IEEEauthorrefmark{3}Faculty of Economic and Social Sciences and Solvay Business School, Vrije Universiteit Brussel, Belgium, \\ Email: wouter.verbeke@vub.ac.be}
\IEEEauthorblockA{\IEEEauthorrefmark{4}Grandata Labs, Buenos Aires, Argentina, Email: charles@grandata.com}
\IEEEauthorblockA{\IEEEauthorrefmark{5}Department of Decision Analytics and Risk, University of Southampton, UK}
\IEEEauthorblockA{\IEEEauthorrefmark{6}Faculty of Economics and Business, KU Leuven, Belgium, Email: \{bart.baesens, jan.vanthienen\}@kuleuven.be}
}
\maketitle

\begin{abstract}
Relational learning in networked data has been shown to be effective in a number of studies.
Relational learners, composed of relational classifiers and collective inference methods, enable the inference of nodes in a network given the existence and strength of links to other nodes.
These methods have been adapted to predict customer churn in telecommunication companies showing that incorporating them may give more accurate predictions.
In this research, the performance of a variety of relational learners is compared by applying them to a number of CDR datasets originating from the telecommunication industry, with the goal to rank them as a whole and investigate the effects of relational classifiers and collective inference methods separately.
Our results show that collective inference methods do not improve the performance of relational classifiers and the best performing relational classifier is the network-only link-based classifier, which builds a logistic model using link-based measures for the nodes in the network. 
\end{abstract}

\section{Introduction}\label{intro}
Customer churn prediction (CCP) in the telecommunication industry (telco) has been intensively researched during the last decade.
The market is competitive for the companies since, while it is relatively easy for unhappy customers to change providers, acquiring new customers is more expensive than retaining current ones and happy customers are more likely to attract others \cite{berson2002building,verbeke2011building}.
In addition, the companies gather an abundance of data about their customers, such as demographic information, usage behaviour and call detail records (CDR) which, when analysed in the correct way, can provide valuable information and enhance the likelihood of the company to thrive in this fiercely competitive market.

In the last years, various studies have confirmed that incorporating network effects in CCP models can improve their performance greatly.
One way to model network effects, is by applying propagation algorithms to call networks and thus simulate how churners might influence other users in the network.
The propagation algorithms produce a score for each customer, that can either be used as an attribute in classical binary classifiers or seen as that person's probability of churning.
The call networks are built by aggregating CDR datasets to create a social network of customers who are connected if they have a relationship, which in this case means that they have used their cell phones to connect.
Based on the network, information flow between the customers through the strength of the connections can be simulated and used to make inferences about the customer's characteristics, such as the propensity to churn.
This is called network learning \cite{macskassy2007classification}. 
Various propagation methods for network learning exist.
Together we refer to them as Relational Learners (RL), since they are used to learn from relationships in a network.
They can be separated into two groups based one their purpose: Relational Classifiers (RC) and Collective Inference methods (CI).  
\IEEEpubidadjcol
In this study we expand on NetKit, a relational learning framework, presented by Macskassy and Provost \cite{macskassy2007classification} which was later adapted for predicting churn in telcos \cite{verbeke2014social}.
In the latter study, four relational classifiers and five collective inference methods were applied to call networks to investigate different ways of incorporating relational learners and whether they improve predictions.
Here we focus on one of those ways, namely using the relational learners as churn prediction models themselves, thus interpreting the scores as probabilities of churn.
We have collected seven distinct CDR datasets from across the world, to which we applied 24 different combinations of RC and CI.
The goal of the study is threefold.
Firstly, we would like to know if there is a relational learner which outperforms the others.
Secondly, by looking at the relational classifiers themselves, we seek to discover if they have an internal ranking in terms of performance.
Finally, we investigate the same for the collective inference methods to answer the ultimate questions of whether combining them with relational classifiers improves the performance of customer churn prediction models in telco.

Our contributions are the following:
\begin{itemize}
\item
There is a group of relational learners which outperform the rest. Most of them use the network-only link-based classifier.
\item
The worst performing learners are most often the same for all performance measures. These are the learners that apply the iterative classification collective inference method.
\item
In the case of customer churn prediction in telco, there is no added benefit of using collective inference methods together with relational classifiers, which perform better when used on their own.
\end{itemize} 

The rest of this paper is organized as follows.
In section \ref{litterature} we present related work regarding churn prediction in telco as well as inferencing in networked data.
After that, we describe the methodology used in this research. 
We start with a discussion about networked data followed by a description of the algorithms used in the experiments.
Next, we provide a short description of the measures used to evaluate the performance of the relational learners.
We briefly describe the datasets we used and then explain our experimental setup.
Subsequently we present the results of our experiments.
Finally, the paper concludes with future work.

\section{Related Work}\label{litterature}
As a research area, customer churn prediction modelling is already well established.
It is a classification problem which has been studied intensely in various domains where the relationship with current customers is seen as a valuable asset for the company. 
Customer churn prediction has been applied in the banking sector \cite{xie2009customer, lariviere2004investigating,glady2009modeling}, by insurance companies \cite{guillen2012time,gunther2014modelling}, internet service providers \cite{khan2010applying} and in the telecommunication industry, which is the industry on which we focus here.
We refer to \cite{verbeke2012new} for an overview of commonly used classification techniques for churn prediction in telco and a benchmarking study of those techniques.

The benchmarking study \cite{verbeke2012new} as well as many other papers mostly rely on local variables, meaning that no social behaviour was incorporated or they are not explicit about social effects in their models.
However, social variables, such as social interaction and calling behaviour, are important factors when it comes to churn in telco. 
In this context, \textit{local variables} include demographic and usage features for the entities in the dataset, without any information about connections to other entities.
In contrast, \textit{social variables} contain some information about the links to other entities \cite{baesens2014analytics}.
In recent studies where social effects have been incorporated in the models, it has been shown that the model performance improves greatly.
In some studies the datasets have been enriched with network variables, such as degree, transitivity and centrality, before building a model using binary classifiers \cite{zhang2012predicting,modani2013cdr,backiel2014mining,benedek2014importance}.
Including these kinds of variables in the dataset adds valuable information that is different from the local variables, and thus better models are produced.

Other studies exploit some kind of propagation algorithms that spread 'churn influence' throughout the network to simulate how churners might possibly affect the people they are connected with.
The spreading activation algorithm  \cite{dasgupta2008social}, which is often compared to the word-of-mouth effect, has successfully been used for this purpose both on its own \cite{dasgupta2008social,backiel2015combining} and to produce scores that are then used as variables in non-relational classifiers, such as logistic regression and decision trees \cite{kusuma2013combining,kim2014improved}.
Lu and Getoor \cite{lu2003link} introduced a method called the network-only link-based classifier, which learns a logistic regression model for the nodes in the network using link-based network variables. 
In other words, each node is classified using only information about connections to other nodes.

In addition, these two modelling approaches have been combined in various ways with one study showing that the best combination model is achieved when a binary classifier is used to build a new model, using the scores resulting from non-relational classifiers and relational learners (in this case spreading activation) as variables \cite{backiel2015combining}. 

The increase in predictive power of models that take into account social network effects can partly be explained by homophily and social influence.
Homophily is used to describe peoples' tendency to associate with those that are similar to them \cite{mcpherson2001birds}.  
Studies have shown that there is significant difference in the phone usage of different genders and age groups, and the usage patterns can be used to predict the demographic features \cite{kovanen2011temporal,sarraute2014study}.
Additionally, homophily can be used to predict similarities between people who interact frequently or to predict interactions between people who behave in a similar way \cite{zhang2012predicting,rhodes2009inferring}.
Social influence, on the other hand, happens when a person's behaviour is influenced by others around them \cite{shalizi2011homophily}.

Finally, in addition to CCP in telco, social network analysis has also been shown to give improved results in other areas, such as social network online games \cite{liao2015customer} and social security and credit card fraud \cite{van2014gotcha, van2015apate}.

\section{Methodology}
\subsection{Networks}
Networks consist of nodes, which may represent real life entities such as people, and the connections or relationships between them, the edges \cite{newman2010networks}.
Weighted and labelled networks are often represented by the triple
\begin{equation}
\mathcal{G}=(\mathbf{V}, \mathbf{E}, \mathcal{X})
\end{equation}
where
\begin{equation}
\mathcal{X}=\{c_1,\dots,c_m\}
\end{equation}
is a set of $m$ classes.
The first component in the triple, $\mathbf{V}$, consists of a vector of $n\in\mathbb{N}$ vertices $\mathcal{V}=\{v_1,\dots v_n\}$ and a vector of labels $\mathcal{L}=\{l_1,\dots, l_n\}$ where each $l_i\in\mathcal{X}$ is the class of node $v_i$.
Similarly, the edge part of the graph, $\mathbf{E}$, is composed of two elements, edges and weights. 
The edges $\mathcal{E}$, are a set of two-subsets of $\mathcal{V}$, where the edge $e_{ij}\in\mathcal{E}$ means that there exists a connection between nodes $v_i$ and $v_j$.
The second component of $\mathbf{E}$ are the weights, $w_{ij}$, which represent the strength of the connection between two edges.
As such $w_{i,j} \in \mathbb{R}_{+}$ and we further assume that the networks are undirected, that is
\begin{equation}
w_{ij}=w_{ji}, \quad \forall ~ i,j\in\{1,\dots,n\}.
\end{equation}

For the purpose of this study, we suppose that not all of the class labels are known.
Therefore we make a distinction between the set of unknown $\mathcal{V}^U$ and know $\mathcal{V}^K$ nodes.  
\begin{figure*}
\centering
 \hspace*{\fill}%
\subfloat{\includegraphics[scale=0.2]{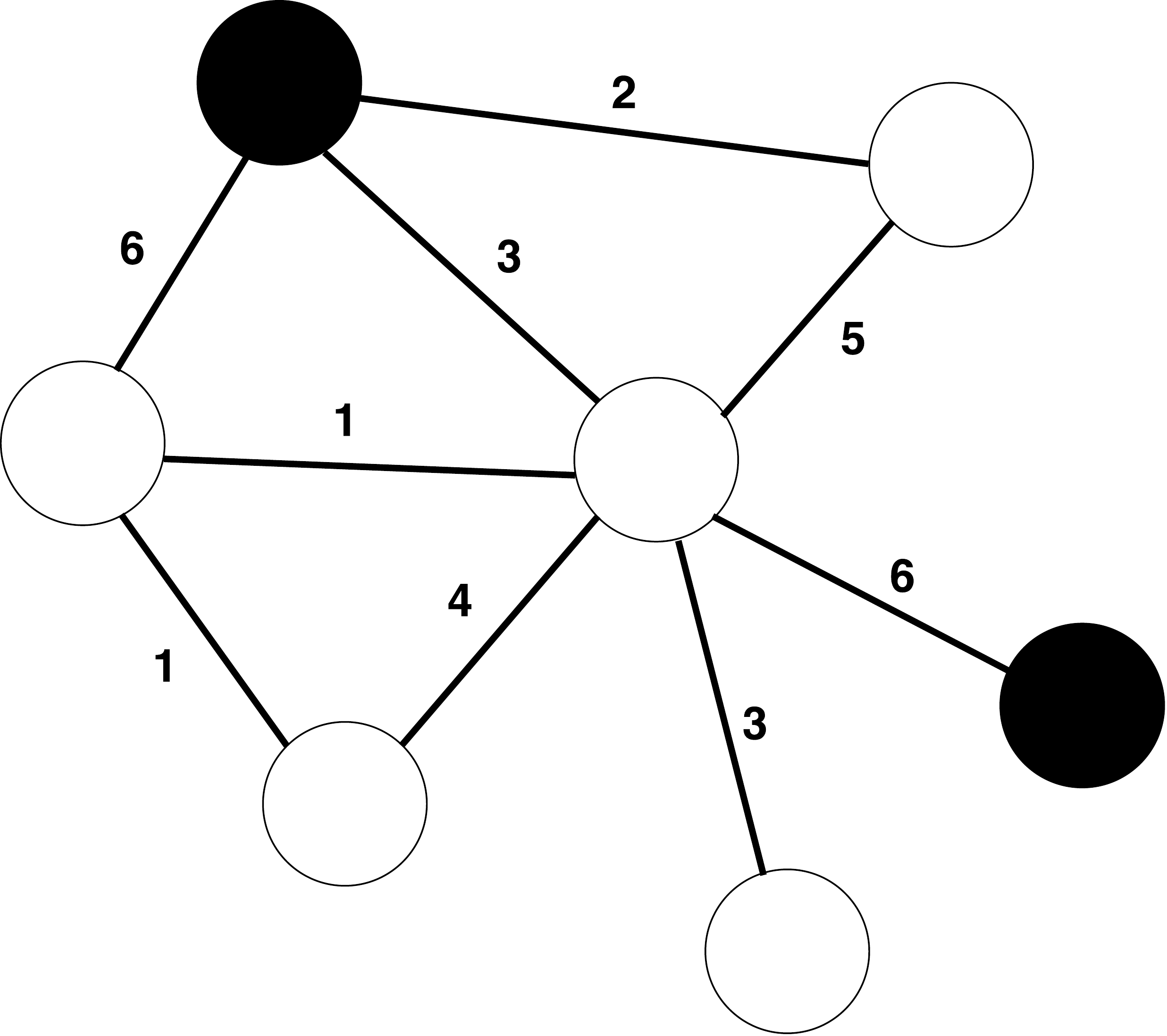}} \hfill
\subfloat{\includegraphics[scale=0.2]{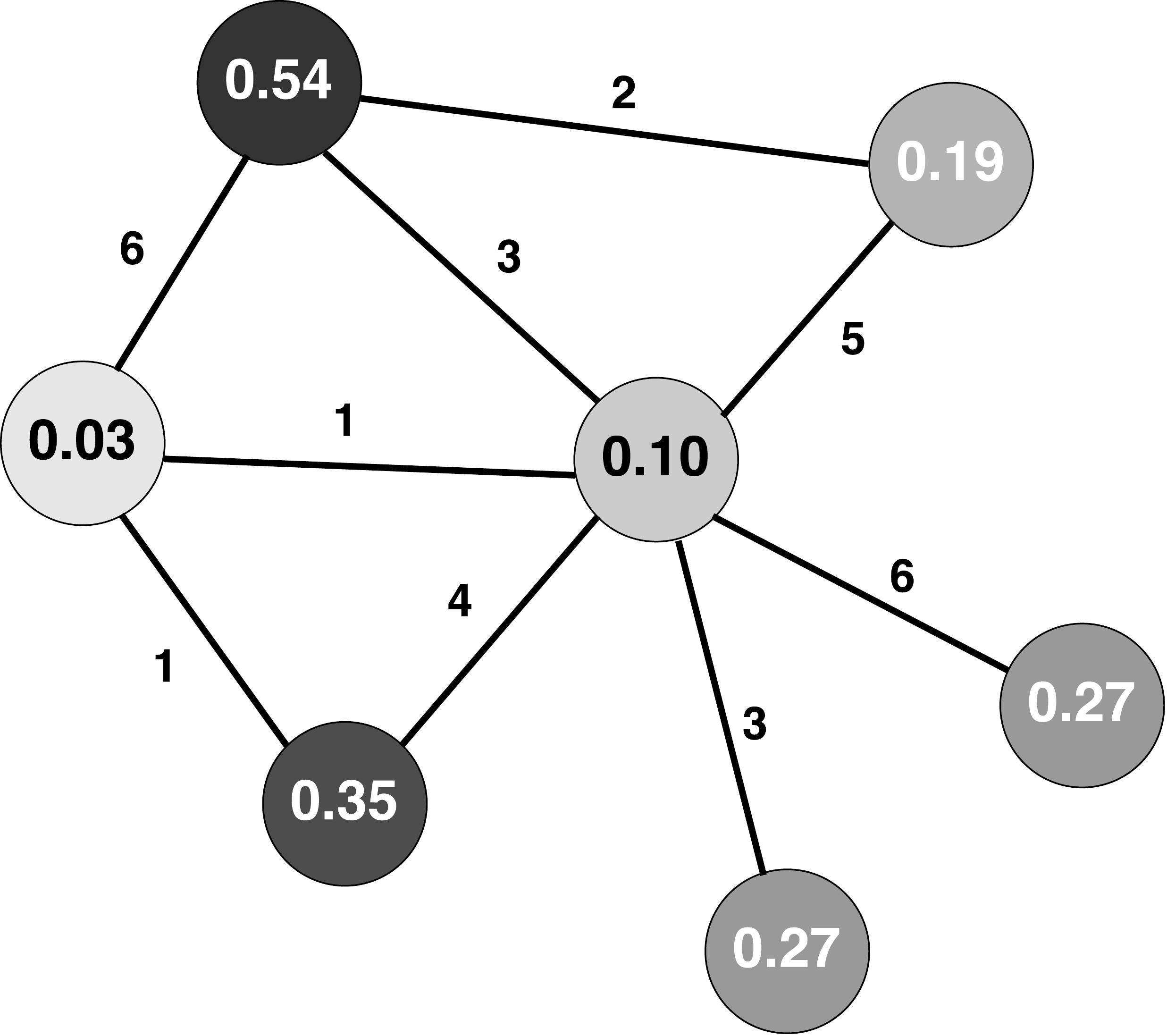}}
 \hspace*{\fill}%
\caption{\label{introRL} The figure shows an example of an application of a relational learner.  The figure on the left, displays a graph with seven customers, of which two have churned (black) and five have not churned (white).  The figure on the right shows the same network after the RL has been applied. Each customer now has a score or probability of churning.}
\end{figure*}
Lastly, we define the neighborhood of a node $v_i$ as being the collection of the node itself and all nodes that are directly linked to it or
\begin{equation}
\mathcal{N}_i=\{v_i\}\cup\{v_j\in\mathcal{V}~|~ w_{ij}\ne 0\}.
\end{equation}
There are various ways to define the strength of links between two nodes in a network.
In the case of call networks, indicating whether or not two customers had a relationship during a specific time period will result in a binary network.
Other, more advanced possibilities, include keeping track of how much time customers spend talking to each other or how often they talk and send messages during a certain period of time.
Representing the weights in different ways will result in different insights.
In addition, the links may have different weights depending on when they were made.
In that way, recent activity between two customers can be given more importance than older activity.
To model this in the network, the weights at time $t$, $w_{ij,t}$ can be exponentially weighted in time by 
\begin{equation}
\label{formula}
(w_{ij})_t= e^{-\gamma t}w_{ij,t} 
\end{equation}
where $\gamma$ is the decay constant. 
The final weights are obtained by aggregating all $(w_{ij})_t$ for the whole time period.
Weighted networks have been successfully used in credit card and social insurance fraud detection \cite{van2014gotcha,van2015apate}.

\subsection{Learning in Networks}
Networks representing relationships between people can be exploited for network learning, thus simulating social influence.
In CCP this is particularly interesting since it is possible to estimate how churners affect the non-churners and subsequently how 'churn-influence' spreads through the network.
Relational learners exploit the information flow between interlinked entities in a network.
Assuming we have available information describing the characteristics of the nodes, the links between them can be used to infer these characteristics if they happen to be unknown for some nodes.
The framework and toolkit NetKit was developed with this in mind \cite{macskassy2007classification}. 
It can be applied to a range of networked data, thus combining various methods to make inferences about interlinked nodes with unknown labels in a network.

In their paper, Verbeke et al. \cite{verbeke2014social} adjusted the NetKit framework to fit the requirements of customer churn prediction in telco. 
To make it applicable to the more specific problem setting, a few changes were made.
Firstly, churn prediction is a binary classification problem, meaning that there are only two classes which results in
\begin{equation}
\mathcal{X}=\{c_0,c_1\},
\end{equation} 
where $c_0$ and $c_1$ represent non-churners and churners, respectively.
Secondly, some of the collective inference methods in \cite{macskassy2007classification} were too slow for the large call networks, and were therefore adjusted to run pseudo-simultaneously instead of sequentially.
Finally, Verbeke et al. \cite{verbeke2014social} introduced the aspect of time in CCP to NetKit.
In NetKit, it is always assumed that only some of the nodes have unknown labels and the rest have known labels which are used to infer the unknown ones.
In churn prediction, on the other hand, the aim is to make predictions into the future, where all labels are unknown.
To implement this, Verbeke et al. suggested training the algorithms at a specific time, $t$, where all labels are known, and use the resulting scores as the estimated labels at time $t+1$.
This means that $\mathcal{V}^K=\emptyset$ at time $t+1$ before the analyses are started.
This is depicted in Fig. \ref{introRL} which shows a small call network before and after a relational learner is applied.
The figure on the left shows the network at time $t$ where all labels are known and two people have churned (black nodes).
The figure on the right shows the same network at time $t+1$ when a relational learner has been used to simulated the propagation of 'churn-influence' resulting in a score or churn probability for each node.

In this research, relational learners are split into two groups: Relational Classifiers and Collective Inference methods.
\begin{table}
\centering
\caption{\label{T:RC} Relational Classifiers}
\scalebox{0.95}{
\begin{tabular}{|l|m{6.8cm}|}
\hline
\textbf{Abbreviation}&\textbf{Description}\\
\hline
WVRN&The weighted vote relational neighbor classifier infers a score based on the weighted labels of the connected nodes.\\ \hline
CDRN&The class distribution relational neighbor classifier assigns a label by looking at the distribution of classes of connected nodes \cite{rocchio1971relevance}.\\ \hline
NLB&The network-only link-based classifier determines a score by learning a logistic regression model using the link-based measures of the nodes. \cite{lu2003link}.\\ \hline
SPA RC& The spreading activation relational classifier is the classification part of the spreading activation algorithm.  It computes scores by looking at weights and labels of nodes connected to neighboring nodes \cite{dasgupta2008social}. \\ \hline
\end{tabular}}
\end{table}

\textit{Relational Classifiers} (RC) are the methods which infer class labels for each node in a network based on the strength of links to other nodes and the labels of those nodes.
They perform a single, local operation going from node to node until all have been classified.
Different methods exist.
The ones that were used in this study are discussed in table \ref{T:RC}.

When going through the network in this manner, it is easy to see that the classification might not be very stable.
Once the first node has been classified, its class label is used to infer the class label for the second node, which in turn might change, which could again have an effect on the first node. 
When applying the RC a single time, this effect is not captured.  
To do so, collective inference methods are used to regulate the inferencing process.

\textit{Collective Inference} methods (CI) are procedures which infer class labels for the nodes in a network while taking into account how the inferred labels affect each other.
They decide in which order the nodes are labelled and how a final label is determined.
They have been shown to improve the performance of relational classifiers in genomes and bibliographic networks \cite{sen2008collective,jensen2004collective}.
Table \ref{T:CI} documents the CI used in this study.
In general, the CIs work in a very similar way performing two operations iteratively until some terminating requirement is reached.
First, a relational classifier is applied to each node in the network and then the score of each node is updated using the results of the RC.

When CIs are applied, the resulting scores have a tendency to level out, and as a result, they don't have much variation.
That is, when the methods repeatedly classify the nodes, a smoothing effect of the churn influence occurs, resulting in very little distinction of churners and non churners.
This was verified in the early stages of this study,  showing that for some CIs the variation in scores decreases very quickly.
As a result, early stopping was implemented in all CI algorithms.
The early stopping is similar to the mechanism which was already implemented for the spreading activation method. 
The criterion is that as soon as the variation in the scores is less than some threshold, the inferencing stops.
The threshold was determined after a sensitivity analysis was performed on part of the data.
\begin{table}
\centering
\caption{\label{T:CI} Collective Inference Methods}
\scalebox{0.95}{
\begin{tabular}{|l|m{6.8cm}|}
\hline
\textbf{Abbreviation}&\textbf{Description}\\
\hline
GS&Gibbs sampling is well known in image retrieval.  It is an iterative method which applies an RC to the network, first for a burn-in period without keeping track of class inference and subsequently applying it and adding up the scores until all iterations are finished. In each iteration, a class label is sampled from the resulting vector of the RC. At the end the scores are normalized to get final scores \cite{geman1984stochastic}.\\\hline
IC&Iterative classification iteratively applies a relational classifier to the network, assigning the label which most often occurs in the resulting vector of estimates to each node \cite{lu2003link}.\\\hline
RL&Relaxation labelling simultaneously infers scores for each node in the network by applying a relational classifier in each iteration and using the resulting estimates as input for the next iteration.  \cite{chakrabarti1998enhanced}\\\hline
RL AS&Relaxation labelling with simulated annealing is an extension of RL with an added annealing term. It is related to the PageRank algorithm \cite{langville2011google}.\\\hline
SPA CI& The spreading activation collective inference method is the CI part of the SPA method. It initializes scores the same way as the gibbs sampler and then iteratively applies RC to the network and updates the scores. \cite{dasgupta2008social}.\\
\hline
\end{tabular}}
\end{table}

Each of the four relational classifiers can be combined with one of the five collective inferencing methods or applied on its own, which results in a total of 24 combinations of methods.
\subsection{Performance Measures}
To evaluate the performance of the models, lift, AUC and the H-measure will be used. 
Firstly, we look at the lift measure \cite{hung2006applying}, which represents how much better a model is at identifying churners than one would find in a random sample.
Although lift at $10\%$ is most commonly used we have chosen to use lift at $0.5\%$ and $1\%$ since most of the datasets contain more than a million customers and $10\%$ of customers is already too many for the company to approach in a retention campaign.  
The smaller portions will give a more realistic view of the performance.

Secondly, we will use the well known area under the receiver operating characteristics curve (AUC) which represents in a single number the trade off between specificity and sensitivity of the model \cite{verbraken2013novel}.

Finally, although AUC is well known and widely used it has been shown to be incoherent when comparing different methods.  
Since the goal of this study is to compare methods, we opted to use the H-measure as well. 
The H-measure was introduced as a coherent alternative to AUC, since it compares all methods to the same metric \cite{hand2009measuring}.
In contrast to the AUC, the H-measure takes into account misclassification costs and in fact seeks to minimize them, to realize a value for performance.

\section{Experimental Set Up}\begin{table}
\caption{\label{T:datasets} Descriptions of datasets}
\scalebox{0.9}{
\begin{tabular}{|m{0.4cm}|m{1cm}|m{0.5cm}|m{1.5cm}|m{0.65cm}|m{1.3cm}|m{1cm}|}
\hline
\textbf{ID}&\textbf{Origin}&\textbf{Year}&$\#$\textbf{Customers}&\textbf{Churn Rate}&\textbf{Sparsity}&\textbf{Contract Type}\\ \hline 
\smallskip
BC1&Belgium&2010 &1.41 Million&$4.4\%$&$7.93\cdot 10^{-7}$&Prepaid\\
BC2&Belgium&2010&1.21 Million&$0.84\%$&$2.20\cdot 10^{-6}$&Postpaid\\
GD1&North America&2015&1.57 Million&$0.71\%$&$3.14\cdot 10^{-6}$&Postpaid\\
GD2&North America&2015&1.32 Million&$2.5\%$&$1.69\cdot 10^{-6}$&Prepaid\\
BP1&Europe&2008 &4.33 Million&$8.5\%$&$9.42\cdot 10^{-7}$&Unknown\\
BP2&Europe&2008&4.52 Million&$3.5\%$&$9.44\cdot 10^{-7}$&Unknown\\
IS&Iceland&2015&93 Thousand&$2.2\%$&$1.04\cdot 10^{-4}$&Postpaid\\ \hline
\end{tabular}}
\end{table}

For the experiments we have collected and analysed seven CDR datasets from around the world.
The datasets together with some of their characteristics can be seen in table \ref{T:datasets}.
As the table shows, the number of customers in a network varies from a hundred thousand to over four million and the churn rates, displayed for the prediction month, are also quite varied, from under a percent to more than eight percent.
\begin{figure*}[!b]
\centering
\includegraphics[scale=0.45]{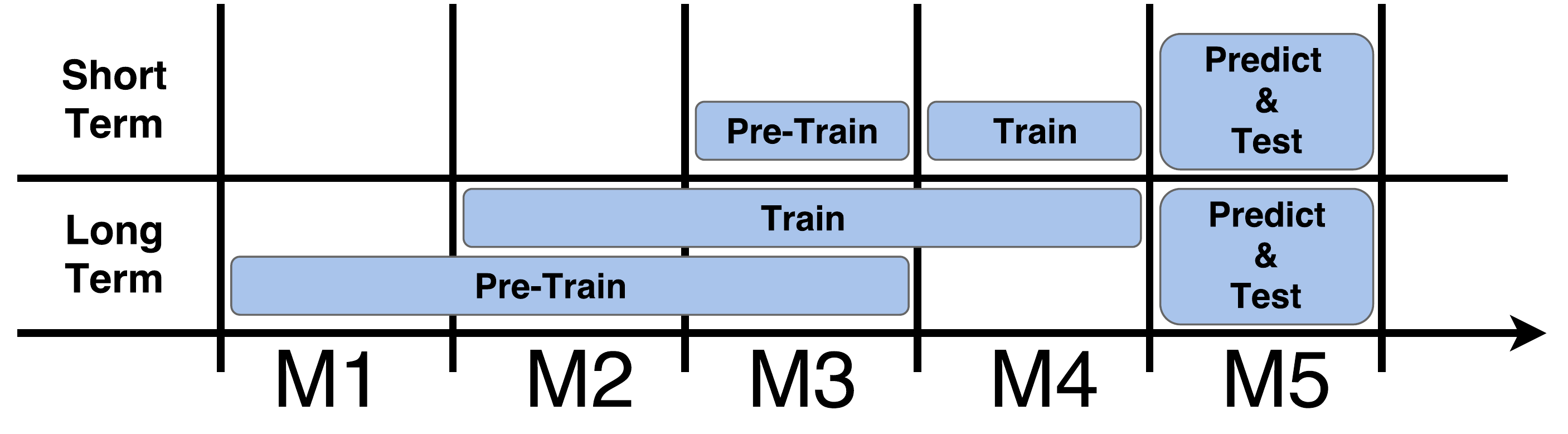}
\caption{\label{timeline} The figure shows how the datasets were split up to build the networks for pre-training, training and predicting in the short and long term setting.}
\end{figure*}
The table also shows the sparsity, or the fraction of non-zero elements, of the networks and thereby gives and indication of how connected the customers are to each other. 
The contract types are both prepaid and postpaid.
Clearly, there is great variety in the datasets and their origin, which is beneficial when it comes to drawing general conclusions from our experiments.

All the datasets were preprocessed in the same way.
Only users of the respective provider were considered and thus only within network correspondence.
In addition, text messages were not taken into account, only phone calls and -- based on an exploratory study --  all phone calls lasting less than four seconds were disregarded, because they do not reflect the connection behaviour that we are modelling within the network.

As the CDR records for all of the datasets span six months, the experiment was set up to make the most of all that data.
Based on previous studies \cite{backiel2014mining,verbeke2014social} and expert knowledge, churn was defined as being inactive for 30 days and then the churn day was defined as the day on which the customer became inactive.
This also ensures the consistency between datasets, since  actual churn dates were only available for a couple of them. 
Because of this definition, the last month can not be used for the analysis but only for building churn labels.
This leaves five months of data to build the models and predict churn, which we name M1, M2, M3, M4 and M5, see Fig. \ref{timeline}.

\begin{figure*}
\centering
\includegraphics[scale=0.85,clip=true, trim=0cm 1.2cm 0cm 0cm]{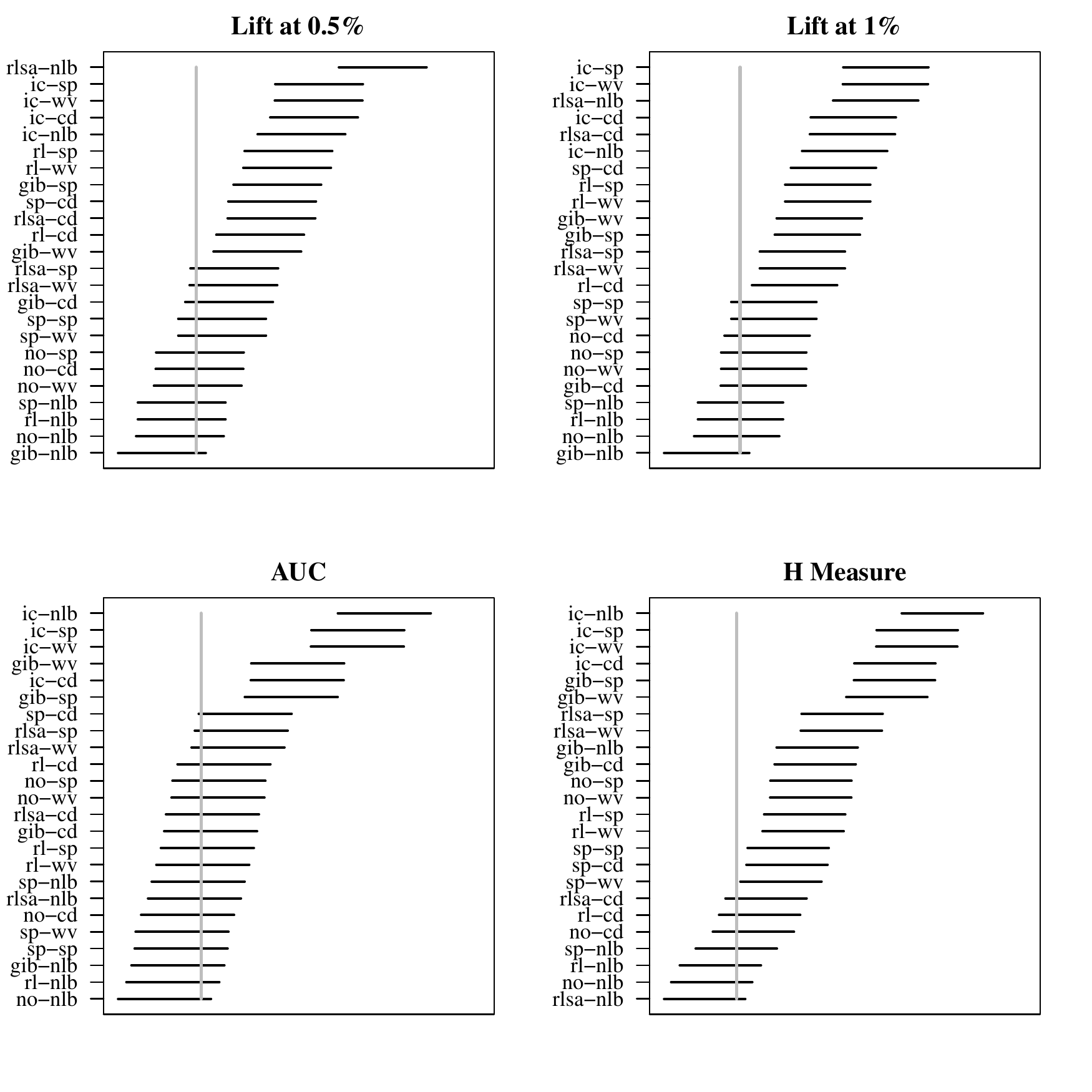}
\caption{\label{nemenyi}The figures show the comparison of all methods against each other using the post-hoc Nemenyi test for the four performance measures. The best method is displayed in the bottom left corner of each figure with the methods that performed significantly worse on the right side of the vertical line which represents the $95\%$ confidence level of the best performing method. }
\end{figure*}
We define two types of churn periods: long term and short term.
The long term period is three months and the short term is one month.  
Two of the relational classifiers, CDRN and NLB, need to be pre-trained on a previous time frame to create reference vectors on which the classification is based.
Therefore, a pre-train time frame was defined to have the same length as the train period but starting one month earlier.
As a result, in the long term setting months M1, M2 and M3 are used for pre-training, months M2, M3 and M4 are used for training and in the short term setting, month M3 is used for pre training and month M4 is used for training.  In both cases the prediction month is M5. See Fig. \ref{timeline}.

In addition, the networks were built using two types of edges, number of phone calls and length of phone calls.
In both cases, the edges were weighted in time with decay as in equation \ref{formula}.

Because of this setup, there were four different networks for each dataset to which each of the 24 relational learners were applied to.
This resulted in 28 scores for each method.
The scores from each of the relational learners were subsequently compared to the churn labels in month M5 to evaluate the models using the four performance measures, $0.5\%$ lift, $1\%$ lift, AUC and $H$-measure.
Finally, various tests were applied to investigate the statistical significance of the results.
\section{Results}
We will present the results to each of the research objectives posed in section \ref{intro}.
We follow the guidelines for statistical comparisons of classifiers over multiple datasets presented by \cite{demvsar2006statistical}.
To evaluate the statistical significance of the comparisons, we start by performing a non-parametric Friedman test, which tests the hypothesis that the average rank of all the methods is equal.
Only when this hypothesis is rejected is it possible to continue with subsequent analyses to compare the individual methods.
Here, we apply the post-hoc Nemenyi test \cite{nemenyi1962distribution} to compare the differences between all methods and to discover which differences are significant.
\begin{figure*}
\centering
\subfloat[Differences of Relational Classifiers]{\includegraphics[scale=0.4]{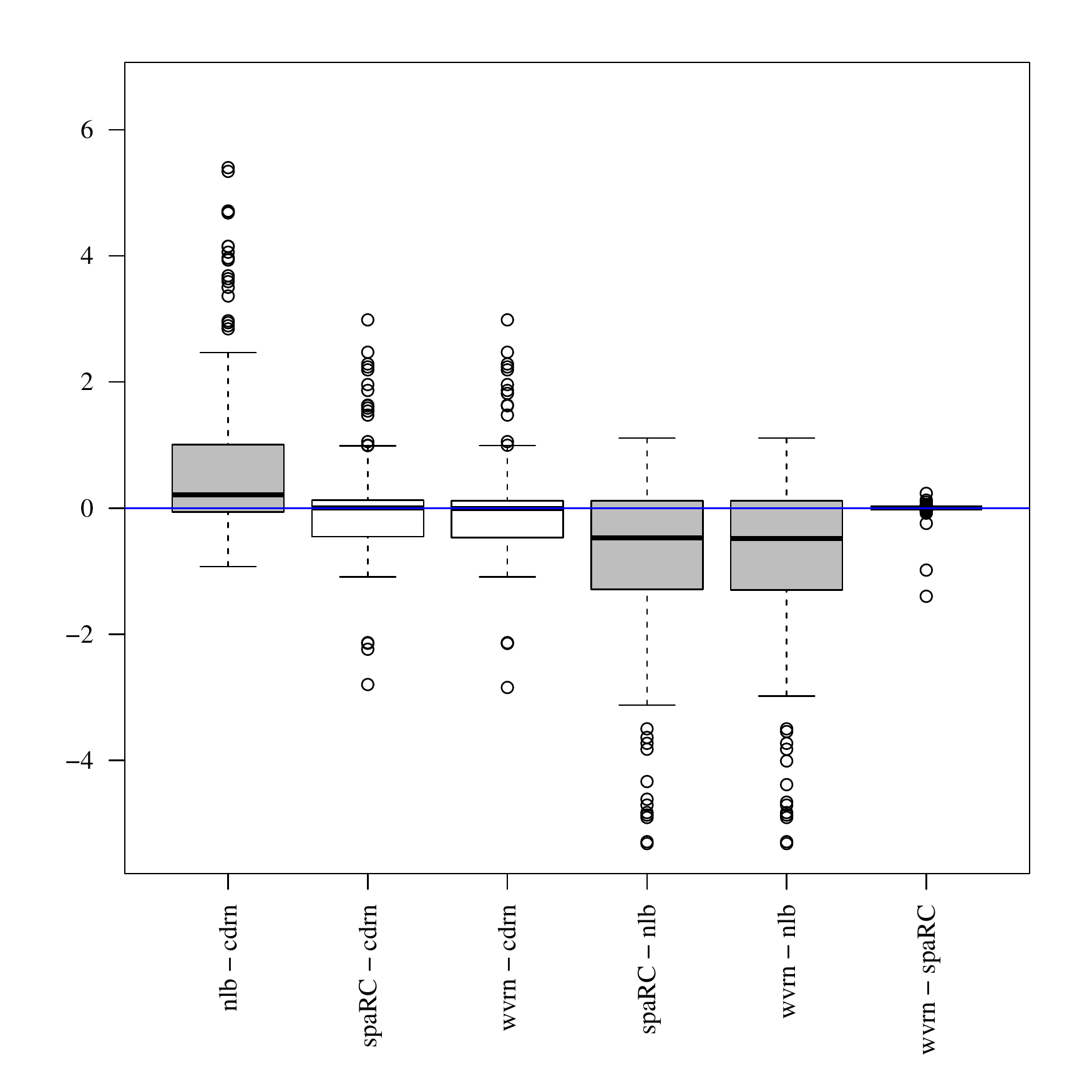}
\label{diffRC}}
\subfloat[Differences of Collective Inference Methods]{\includegraphics[scale=0.4]{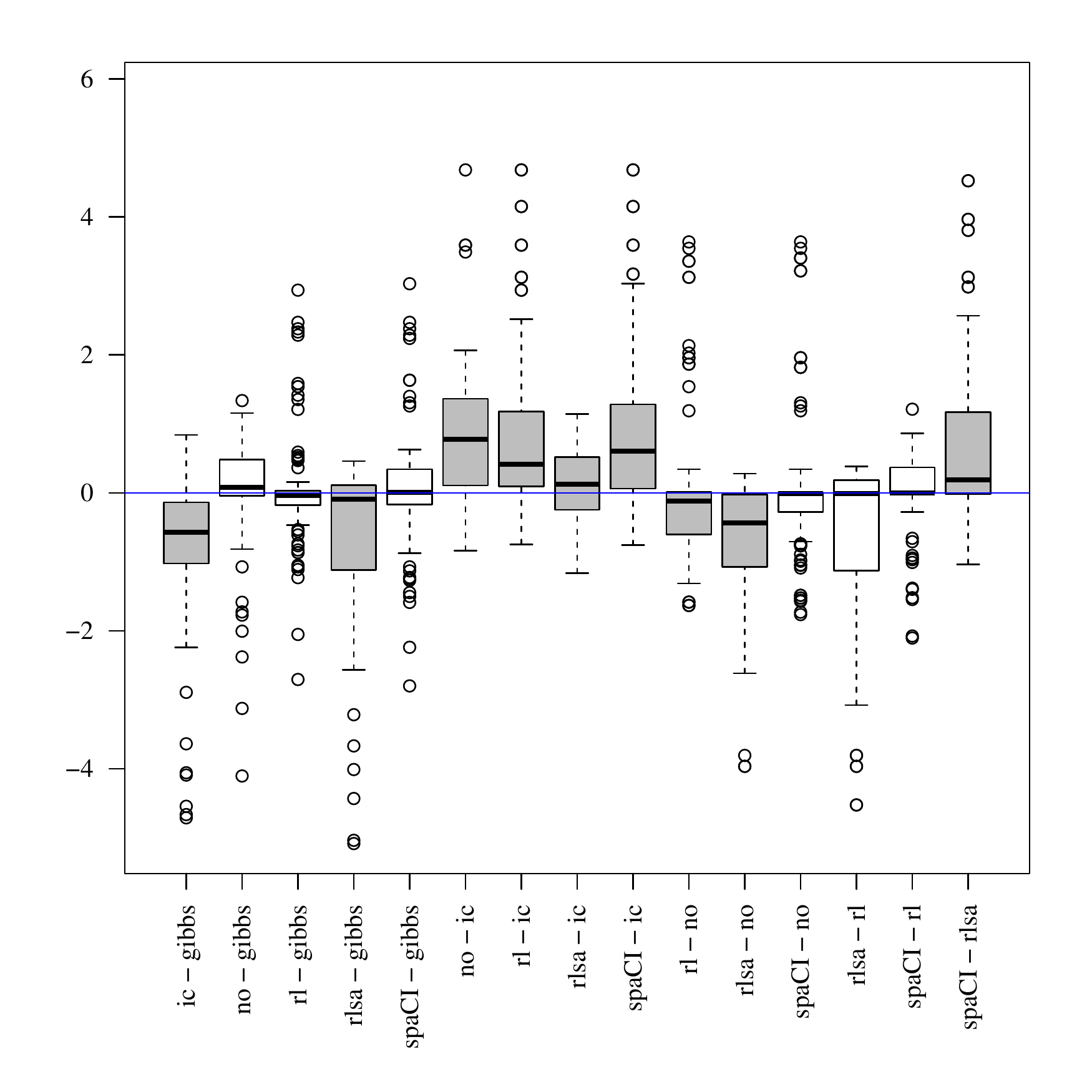}
\label{DiffCI} }
\caption{The figure shows boxplots of the differences in performance of the relational classifiers and the collective inference methods.  Grey colored boxes represent statistically significant differences and the white boxes not statistically significant differences, at the $95\%$ confidence level.}
\end{figure*}
\subsection{Comparison of Relational Learners}
First we compare the performance of the 24 relational learners.
To test for a significant difference between them, a Friedman test was applied to the rankings of each of the four performance measures. 
In all cases, the resulting p-value was close to zero, meaning that not all the methods are the same, and at least one of them performs significantly different from the rest.

Since the null hypothesis was rejected we continued by performing a post-hoc Nemenyi test  to compare the differences between all methods.  
The results for the performance measures $0.5\%$ lift, $1\%$ lift, AUC and H-Measure can be seen in Fig. \ref{nemenyi}.  
In each of the subfigures, the line corresponding to each method represents its average rank.
The higher the average rank, the further to the left this line is, with the left endpoint being the average rank itself and the right endpoint the critical difference from the average rank.
The vertical line represents the critical difference at the $95\%$ confidence level of the best performing method.
This means that if another line is situated fully to the right of the vertical line, it performs significantly worse than the best performing method at the $95\%$ confidence level.
We see from Fig. \ref{nemenyi} that for $0.5\%$ lift twelve of the methods are significantly worse than the best one, fourteen for $1\%$ lift, six for AUC and seventeen for the H-measure.
Noticeably, the NLB performs better than the other relational classifiers and methods without a CI perform better as well.

\subsection{Comparison of Relational Classifiers}
In order to answer the second question regarding the best performing relational classifier, we apply a Friedman test.
This time, there are only four different methods, corresponding to the four classifiers.  
For all four measures the resulting p-value was less than 0.01, which means that there is some difference between the classifiers. 
Therefore, we proceed to compare them in a post-hoc Nemenyi test.
The results can be seen in Fig. \ref{diffRC} which shows a boxplot of the differences in performance in $1\%$ lift of each pair of the two classifiers.
The grey boxes mean that the difference is significant at the $95\%$ confidence level whereas the white colored boxes mean there is not a significant difference.
As can be seen  in the figure, the difference of NLB and any other classifier is always significant, which means that NLB performs better than the others classifiers.

\subsection{Comparison of Collective Inference Methods}
The last step is to evaluate the differences of the collective inference methods.
The Friedman test applied with the six collective inference methods gave a p-value of less than 0.001 for all of the four measures.
The results of the subsequent post-hoc Nemenyi test on the differences, can be seen in Fig. \ref{DiffCI}.
It is evident from this figure that the IC method always performs significantly worse and not applying a collective inference method is often significantly better.

We conclude by investigating the impact of the collective inference methods in improving the performance of relational classifiers.
The non-parametric Friedman test requires a balanced experimental design without repeated measurements and is therefore is not applicable in this case, since there are 20 methods with CI but only four without.
Instead the non-parametric Kruskal-Wallis test is used.
The null hypothesis, of the two samples originating from the same distribution, is rejected with a p-value of less than 0.01.
The result is that methods without CI perform better than methods with CI.

\section{Conclusion}
\subsection{Main Findings}
In this study, we have made a statistical comparison of the performance of two dozen relational learners when predicting churn in telco.
In addition, we have compared the effects of the two components of the relational learners, namely relational classifiers and collective inference methods.
Applying all these methods to 28 networks from seven datasets, we were able to receive robust results about the significance of the methods.

Firstly, we observed in the comparison of the relational learners that the same group of about a dozen learners consistently performs better than the rest.
Our tests showed a clear separation between the better and the worse methods, measured by all applied performance measures.

Secondly, the statistical superiority of the network-only link-based classifier introduced by Lu and Getoor \cite{lu2003link} was evident.  
This method builds a logistic regression model using link-based features extracted from the network. 
These features represent, for each node, the number of neighbors which have churned and not churned, whether a neighbor has churned and what the most common neighbor property is. 
As such, the classifier captures enough information from the network to accurately predict churn and this proves to be the best approach.

In addition, the iterative classification collective inference method, always performed worse than the other methods, which might be due to the fact that after each step in the iteration labels (0 or 1) instead of scores (ranging form 0 to 1) are produced.
As such, the result is much less refined than for the other CIs.

Finally, we have shown that in telco, but possibly in other applications as well, collective inference methods do not improve the performance of relational classifiers when predicting churn.
Explaining this is not straight forward and requires further research. 
One reason might be that since there are relatively few churn signals in the network, as well as edges, the signals get spread out and are not clear enough afterwards. 
Thus, the  'churn influence' gets too diluted to be meaningful.  
\subsection{Future Work}
There is a whole range of possibilities for future research.
So far, we have only used the scores from relational learners as predictions, but they could also be combined with local and network variables as attributes in models using classical binary classification techniques.
The result would not only be better models, but it would also give a better idea of which learners are the most important.
In addition, the results of relational and non-relational classifiers can be combined in different ways, which could be explored given the high number of datasets that we have.
It would also be beneficial to apply all the methods to other CDR datasets to further support the results and make them more robust.

Another aspect that could be investigated further is how people churn. 
Indeed, the way the networks for the analyses were built, offers the possibility of gaining a better understanding of the churn process itself. 
This could be a very interesting result for telco providers to better understand their clients.

Finally, we would like to mention the relatively novel maximum profit measure \cite{verbraken2013novel,verbeke2012new}.
It is specifically tailored to churn prediction problems and evaluates the models by taking into account the cost and benefits of a retention campaign.
We believe that applying the profit measure would give additional insight and important information when it comes to seeing how much the models are worth.

\bibliographystyle{IEEEtran}
\bibliography{IEEEabrv,benchmark_bib}
\end{document}